\documentclass[a4paper,american,english,pre,twocolumn,floatfix]{revtex4}
\usepackage[T1]{fontenc}
\usepackage[latin1]{inputenc}
\usepackage{amsmath}
\usepackage[dvips]{graphicx}
\usepackage{amssymb}

\makeatletter

\providecommand{\tabularnewline}{\\}

\DeclareMathOperator{\Pre}{Pre}
\DeclareMathOperator{\Post}{Post}
\DeclareMathOperator{\rank}{rank}

\newcommand{\Gershgorin}{Ger\v{s}gorin }
\newcommand{\tauh}{\frac{\tau}{2}}
\newcommand{\equa}[1]{\ref{eq:#1}}

\newcommand{\hD}[2]{h(#1+D_{i,#2} , \eps_{i j_{#2}})=:
  \beta_{i,#2}}
\newcommand{\mysubstack}[2]{\substack{#1\\#2} }
\newcommand{\MyRe}{\mathrm{Re}}

\usepackage{babel}
\makeatother
\begin{document}
\selectlanguage{american}

\newcommand{\p}{\varphi}

\selectlanguage{english}

\newcommand{\eps}{\varepsilon}

\selectlanguage{american}

\newcommand{\epshat}{\hat{\varepsilon}}

\newcommand{\reset}{\rightarrow}

\selectlanguage{english}

\title{Speed of synchronization in complex networks of neural oscillators\\
Analytic results based on Random Matrix Theory}

\author{Marc Timme$^{1-4}$, Theo Geisel$^{1-3}$, and Fred Wolf$^{1-3}$}

\affiliation{$^1$ Max Planck Institute for Dynamics Self-Organization \\
and $^2$ Department of Physics, University of Göttingen, \\
and $^3$ Bernstein Center for Computational Neuroscience, 
37073 Göttingen, Germany; \\
$^4$ Theoretical and Applied Mechanics, Cornell University, Kimball Hall, Ithaca, NY 14853, USA
}

\begin{abstract}
We analyze the dynamics of networks of spiking neural oscillators.
First, we present an exact linear stability theory of the synchronous
state for networks of arbitrary connectivity. For general neuron rise
functions, stability is determined by multiple operators, for which
standard analysis is not suitable. We describe a general non-standard
solution to the multi-operator problem. Subsequently, we derive a
class of rise functions for which all stability operators become degenerate
and standard eigenvalue analysis becomes a suitable tool. Interestingly,
this class is found to consist of networks of leaky integrate and
fire neurons. For random networks of inhibitory integrate-and-fire
neurons, we then develop an analytical approach, based on the theory
of random matrices, to precisely determine the eigenvalue distribution.
This yields the asymptotic relaxation time for perturbations to the
synchronous state which provides the characteristic time scale on
which neurons can coordinate their activity in such networks. For
networks with finite in-degree, i.e. finite number of presynaptic
inputs per neuron, we find a speed limit to coordinating spiking activity:
Even with arbitrarily strong interaction strengths neurons cannot
synchronize faster than at a certain maximal speed determined by the
typical in-degree.
\end{abstract}
\maketitle
\textbf{The individual units of many physical systems, from the planets
of our solar system to the atoms in a solid, typically interact continuously
in time and without significant delay. Thus at every instant of time
such a unit is influenced by the current state of its interaction
partners. Moreover, particles of many-body-systems are often considered
to have very simple lattice topology (as in a crystal) or no prescribed
topology at all (as in an ideal gas). Many important biological systems
are drastically different: their units are interacting by sending
and receiving pulses at discrete instances of time. Furthermore, biological
systems often exhibit significant delays in the couplings and very
complicated topologies of their interaction networks. Examples of
such systems include neurons, which interact by stereotyped electrical
pulses called action potentials or spikes; crickets, which chirp to
communicate acoustically; populations of fireflies that interact by
short light pulses. The combination of pulse-coupling, delays, and
complicated network topology formally makes the dynamical system to
be investigated a high-dimensional, heterogeneous nonlinear hybrid
system with delays. Here we present an exact analysis of aspects of
the dynamics of such networks in the case of simple one-dimensional
nonlinear interacting units. These systems are simple models for the
collective dynamics of recurrent networks of spiking neurons. After
briefly presenting stability results for the synchronous state, we
show how to use the theory of random matrices to analytically predict
the eigenvalue distribution of stability matrices and thus find the
speed of synchronization in terms of dynamical and network parameters.
We find that networks of neural oscillators typically exhibit speed
limits and cannot synchronize faster than a certain bound defined
by the network topology.}

\section{Introduction}

Most neurons in the human central nervous systems communicate by sending
and receiving brief stereotyped electrical pulses, called action potentials
or spikes. Via chemical synaptic connections, these spikes induce
changes in the potential across the membrane of the connected postsynaptic
neurons \cite{Rieke:1999}. Due to this mode of communication, these
neurons interact at discrete instances in time only -- and thus behave
substantially different from the interacting units of many physical
systems. Other important characteristics of neuronal communication
are delayed interactions (due to finite propagation speed of the spikes
along axons, non-zero time needed for chemical processes across the
synapses and signal transmission along the dendrites) and a complex
wiring diagram. As in the example of neurons, many networks of interacting
units are not arranged in regular lattices. Instead, single units
form an intricate network of connections that mediate the interactions.
In addition, these connections are often directed, meaning that a
connection from one unit to another does not imply a connection in
the reverse direction. From a dynamical systems perspective, these
aspects -- discrete interaction times, interaction delays, and non-symmetric,
complicated wiring diagram -- make the theoretical investigation of
the exact spiking dynamics of large neural networks a challenging
task.

Previous research has occasionally explicitely considered interaction
delays in analytical calculations; the complicated topology of neural
networks, however, has received much less attention. As a consequence,
if one wants to uncover the dynamics beyond numerical investigations,
one is often restricted to mean field theoretical arguments or focuses
on globally connected networks or on networks of simple local topology
\cite{Bressloff:1997:2791,vanVreeswijk:1994:313,Ernst:1995:1570,vanVreeswijk:1996:1724,vanVreeswijk:1998:1321,Brunel:1999:1621,vanVreeswijk:2000:5110,Hansel:2001:4175,Timme:2002:154105,Roxin:2005:238103}. 

Here we follow the simple and very useful approach of Mirollo and
Strogatz \cite{Mirollo:1990:1645} to represent the state of a one-dimensional
(neural) oscillator not by its membrane potential, but by a phase
that encodes the time to the next spike in the absence of any interactions.
In the limit of infinitely fast processing of incoming signals (post-synaptic
currents), the nonlinear interactions can then be treated analytically
in an exact manner. Following some previous reports \cite{Ernst:1995:1570,Ernst:1998:2150,Senn:2001:1143,Timme:2002:258701,Timme:2004:074101}
that used the advantages \cite{Timme:2003:377} of the Mirollo Strogatz
idea \cite{Mirollo:1990:1645} we here present an analytical approach
to exactly determine the asymptotic dynamics of spiking neural networks
of complicated topology. We particularly focus on how, and how fast,
neurons can synchronize their spikes, i.e. coordinate their activity
in time in networks of random topology. 

The paper is organized as follows. In Section II we briefly introduce
model networks of pulse-coupled neural oscillators and state the research
question. We are interested in the stability of the synchronous state
and its asymptotic synchronization properties. Section III gives the
details of the derivation of nonlinear stroboscopic maps of perturbed
synchronous states in networks of arbitrary connectivity. We explain
the emergence of piecewise analytic maps where the pieces are determined
by the temporal spiking order of a particular perturbation. This results
in a multiple operator nonlinear stability problem. In Section IV,
we derive first order operators from the stroboscopic maps leading
to a stability operator with multiple piecewise linear parts. Since
standard eigenvalue analysis is not appropriate for such multiple
operator problems, we describe an alternative method to demonstrate
stability in Section V. Section VI shows how degeneracy can be enforced,
i.e. how all multiple linear operators can be made degenerate to one
single stability matrix. It turns out that the oscillator rise functions
that guarantee degeneracy are of integrate-and-fire type. For this
stability problem, standard eigenvalue analysis is suitable. For two
ensembles of random networks, we first study their eigenvalue distributions
(Section VII), analytically predict these distributions by measures
derived from Random Matrix Theory (Section VIII) and compare the results
between numerics and analytics (Section IX). In Section X, we discuss
consequences of the eigenvalue distributions for the speed of synchronization
of neural oscillators. We close in Section XI, where we summarize
the results, discuss some of their consequences and give a brief outlook.

This paper presents new aspects and detailed descriptions of the determination
of the asymptotic synchronization time by Random Matrix Theory. Parts
of the results on stability and speed limits to network synchronization
have been reported in brief in references \cite{Timme:2002:258701}
and \cite{Timme:2004:074101}, respectively. Details of the stability
theory, in particular exact eigenvalue bounds and asymptotic stability
in the multi-operator case, not discussed here, can be found in \cite{Timme:2006:NC}.
For effects on parameter inhomogeneities, leading to close to synchronous
patterns of spikes, we refer the reader to \cite{Denker:2004:074103}.

\section{Model of neural oscillators}

Consider a system of $N$ neural oscillators that interact by sending
and receiving pulses via directed connections. The sets $\Pre(i)$
of presynaptic oscillators having input to an oscillator $i$ define
the network connectivity. The number of inputs \begin{equation}
k_{i}:=|\Pre(i)|\label{eq:ki}\end{equation}
 to every oscillator $i$, called in-degree in graph theory \cite{Chartrand:1996}
is non-zero, $k_{i}\geq1$, and no further restriction on the network
topology is imposed unless otherwise stated. 

The state of an individual oscillator $j$ is represented by a phase-like
variable $\phi_{j}\in(-\infty,1]$ that increases uniformly in time,
\begin{equation}
d\phi_{j}/dt=1\,.\label{eq:phidot1}\end{equation}
 Upon crossing a firing threshold, $\phi_{j}(t_{\textrm{f}})\geq1$,
at time $t_{\textrm{f}}$ an oscillator is instantaneously reset to
zero, $\phi_{j}(t_{\textrm{f}}^{+})=0$, and a pulse is sent. After
a delay time $\tau$ this pulse is received by all oscillators $i$
connected to $j$ (for which $j\in\Pre(i)$) and induces an instantaneous
phase jump 

\[
\phi_{i}((t_{\textrm{f}}+\tau)^{+})=U^{-1}\left(U(\phi_{i}(t_{\textrm{f}}+\tau)+\eps_{ij}\right)\]
Here, $\eps_{ij}\leq0$ are the coupling strengths from $j$ to $i$,
which are taken to be purely inhibitory ($\eps_{ij}<0$ if $j\in\Pre(i)$,
$\eps_{ij}=0$ otherwise) and normalized, \begin{equation}
\sum_{j=1}^{N}\eps_{ij}=\eps\,,\label{eq:epsnorm}\end{equation}
throughout this paper. 

The rise function $U$, which mediates the interactions, is monotonic
increasing, $U'>0$, concave (down), $U''<0$, and represents the
subthreshold dynamics of individual oscillators. This models the dynamics
of the membrane potential of a biological neuron that is driven by
a current. Note that the function $U$ need to be defined on the entire
range of accessible phase values. In particular, inhibitory coupling
can lead to negative phase values $\phi_{i}<0$. 

Large sparsely connected networks of inhibitory neurons were known
before to exhibit irregular asynchronous spiking states in which excitatory
drive and inhibitory feedback balance out and fluctuation induce spikes
\cite{vanVreeswijk:1996:1724,vanVreeswijk:1998:1321,Brunel:1999:1621}.
\begin{figure}[htp!]
\begin{center}\includegraphics[%
  width=80mm]{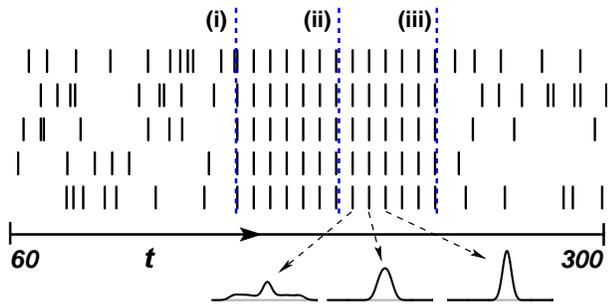}\end{center}

\caption{Irregular, balanced activity coexists with regular, synchronous activity.
This enables switching by external stimulus signals. Random network
of connection probability $p=0.2$ ($N=400$, $I=4.0$, $\eps=16.0$,
$\tau=0.14$). Firing times of five oscillators are shown in a time
window $\Delta t=240$. Vertical dashed lines mark external perturbations:
(i) large excitatory pulses lead to synchronous state, (ii) a small
random perturbation ($|\Delta\phi_{i}|\leq0.18$) is restored (iii)
a sufficiently large random perturbation ($|\Delta\phi_{i}|\leq0.36$)
leads to an irregular state. Bottom: Time evolution of the spread
of the spike times after perturbation (ii), total length $\Delta t=0.25$
each. Decreasing width of the distribution indicates resynchronization.
\label{fig:switch}}
\end{figure}
 However, in a previous study \cite{Timme:2002:258701} we found that
regular states, in the homogeneous case defined by exact spike synchrony,
coexist with irregular states in these networks at the same parameters
(Fig. \ref{fig:switch}). This means that by external perturbations
one can switch between regular and irregular activity. In particular,
strong excitatory synchronous inputs can synchronize the network activity.
Strong random inputs can switch the network back to the balanced state.
If random inputs to the synchronous state are not too strong, the
activity relaxes back to the synchronous state. Two major questions
intrigued us: 1) Why, given an irregular topology of the network,
can the regular synchronous state be stable such that neurons resynchronize
their spikes upon sufficiently small perturbations? 2) What is the
typical time scale for re-synchronization, i.e. how fast can neurons
coordinate their spiking activity if they are not directly interconnected
but interact on large networks of complex topology? 

We address these questions analytically in the following, focusing
on the speed of synchronization. All results are derived for the simplest
of all regular states, the synchronous periodic state, in which all
neural oscillators exhibit identical dynamics. However, a similar
approach can be used for cluster states in which two or more groups
of synchronized oscillators exist \cite{Ernst:1995:1570,Ernst:1998:2150},
as well as for any periodic solution because they can be tracked analytically
in the model system used. In the presence of inhomogeneity, the approach
needs to be modified but similar principles are expected to apply. 

\newpage
\section{Nonlinear Stroboscopic Maps:\protect \\
Emergence of Multiple operators }

The synchronous state 

\begin{equation}
\phi_{i}(t)=\phi_{0}(t)\quad\textrm{for all }i\,,\end{equation}
in which all oscillators display identical phases $\phi_{0}(t)$ on
a periodic orbit such that $\phi_{0}(t+T)=\phi_{0}(t)$, is one of
the simplest states a network of neural oscillators may assume. The
normalization of the coupling strengths (\ref{eq:epsnorm}) ensures
that it exists but does not tell whether or not it is stable and an
attractor of the system. To uncover this, we perform a stability analysis
of the synchronous state the period of which is given by \begin{equation}
T=\tau+1-\alpha\label{eq:period}\end{equation}
 where \begin{equation}
\alpha=U^{-1}(U(\tau)+\eps).\label{eq:alpha}\end{equation}
For inhibitory coupling ($\eps<0$) and sufficiently small delay $\tau<1$
the total input is subthreshold, $U(\tau)+\eps<1$ such that $\alpha<1$.
A perturbation \begin{equation}
\boldsymbol{\delta}(0)=:\boldsymbol{\delta}=(\delta_{1},\ldots,\delta_{N})\end{equation}
 to the phases is defined by \begin{equation}
\delta_{i}=\phi_{i}(0)-\phi_{0}(0)\,.\label{eq:define_delta}\end{equation}
 If we assume that the perturbation is small, in the sense that \begin{equation}
\max_{i}\delta_{i}-\min_{i}\delta_{i}<\tau\label{eq:taubounds}\end{equation}
it can be considered to affect the phases of the oscillators at some
time just after all signals have been received, i.e.~after a time
$t>t_{0}+\tau$ if all oscillators have fired at $t=t_{0}$. Such
a perturbation will affect the time of the next firing events because
the larger the perturbed phase of an oscillator is, the earlier this
oscillator reaches threshold and sends a signal. 

To construct a stroboscopic period-$T$ map, $\boldsymbol{\delta}$
is ordered according to the rank order $\rank(\boldsymbol{\delta})$
of the $\delta_{i}$: For each oscillator $i$ we label the perturbations
$\delta_{j}$ of its presynaptic oscillators $j\in\Pre(i)$ according
to their size\begin{equation}
\Delta_{i,1}\geq\Delta_{i,2}\geq\ldots\geq\Delta_{i,k_{i}}\label{eq:spikeorder}\end{equation}
where $k_{i}$ is the number of its presynaptic oscillators (\ref{eq:ki}).
The index $n\in\{1,\ldots,k_{i}\}$ counts the signals that arrive
successively. Thus, if $j_{n}\equiv j_{n}(i)\in\Pre(i)$ labels the
presynaptic oscillator from which $i$ receives its $n^{\mathrm{th}}$
signal during the period considered, we have \begin{equation}
\Delta_{i,n}=\delta_{j_{n}(i)}\,.\label{eq:Deltaindeltajn}\end{equation}
In addition, we define \begin{equation}
\Delta_{i,0}=\delta_{i}\,.\label{eq:Deltai0}\end{equation}

\begin{figure*}[htp!]
\begin{center}\includegraphics[%
  width=14cm]{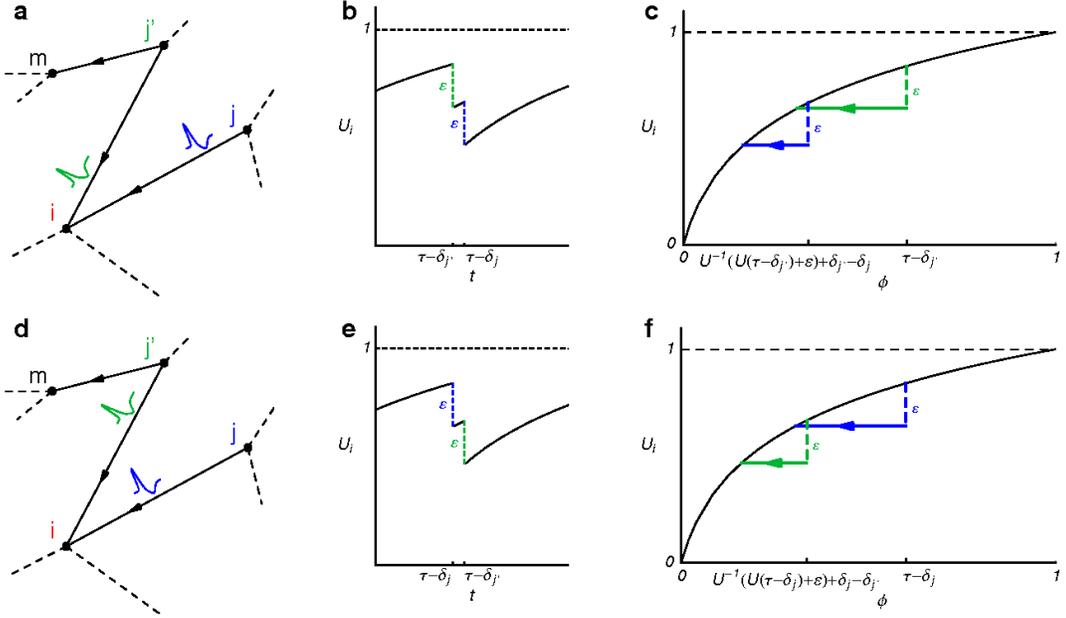}\end{center}

\caption{Two signals arriving almost simultaneously induce different phase
changes, depending on their rank order. The figure illustrates a simple
case where $\Pre(i)=\{ j,j'\}$ and $\delta_{i}=0$, (a)--(c) for
$\delta_{j'}>\delta_{j}$ and (d)--(f) for $\delta_{j}>\delta_{j'\,}$.
(a), (d) Local patch of the network displaying the reception times
of signals that are received by oscillator $i$. Whereas in (a) the
signal from $j'$ arrives before the signal of $j$, the situation
in (d) is reversed. (b), (e) Identical coupling strengths induce identical
jumps of the \textit{potential} $U$ but (c),(f) the \textit{phase}
jumps these signals induce are different and depend on the order of
the incoming signals. For small $|\delta_{i}|\ll1$, individual phase
jumps are encoded by the $p_{i,n}$ , see (\ref{eq:fractions}). The
Figure displays an example for inhibitory (negative, phase-retarding)
coupling but the mechanism generating multiple operators does not
depend on the signs of the coupling strengths. \label{fig:flip_spikes}}
\end{figure*}

For illustration, let us consider an oscillator $i$ that has exactly
two presynaptic oscillators $j$ and $j'$ such that $\Pre(i)=\{ j,j'\}$
and $k_{i}=2$ (Fig.~\ref{fig:flip_spikes}a,d).%
 For certain perturbations, oscillator $i$ first receives a signal
from oscillator $j'$ and slightly later from oscillator $j$. This
determines the rank order, $\delta_{j'}>\delta_{j}$, and hence $\Delta_{i,1}=\delta_{j'}$
and $\Delta_{i,2}=\delta_{j}$ (Fig.~\ref{fig:flip_spikes}a). Perturbations
with the opposite rank order, $\delta_{j}>\delta_{j'}$, lead to the
opposite labeling, $\Delta_{i,1}=\delta_{j}$ and $\Delta_{i,2}=\delta_{j'}$
(Fig.~\ref{fig:flip_spikes}d). In general, relabeling cannot be
achieved by permuting the indices because one oscillator $j'$ may
receive an input connection from yet another one $m$ whereas oscillator
$j$ may not receive this connection.

We now consider a fixed arbitrary perturbation, the rank order of
which determines the $\Delta_{i,n}$ according to the inequalities
(\ref{eq:spikeorder}). Using the phase shift function $h(\phi,\eps)=U^{-1}(U(\phi)+\eps)$
and denoting \begin{equation}
D_{i,n}:=\Delta_{i,n-1}-\Delta_{i,n}\label{eq:D_i,n}\end{equation}
 for $n\in\{1,\ldots,k_{i}\}$ we calculate the time evolution of
phase-perturbations $\delta_{i}$ satisfying the bound (\ref{eq:taubounds}),
starting near $\phi_{0}(0)=\tau/2$ without loss of generality. The
stroboscopic time-$T$ map of the perturbations, $\delta_{i}\mapsto\delta_{i}(T)$,
is obtained from the scheme given in Table \ref{Tab:complexmap}.%
\begin{table}[htp!]
\begin{center}\begin{tabular}{|c|c|}
$t$&
$\phi_{i}(t)$\tabularnewline
\hline 
$\begin{array}{c}
0\\
\tauh-\Delta_{i,1}\\
\tauh-\Delta_{i,2}\\
\vdots\\
\tauh-\Delta_{i,k_{i}}\\
\tauh-\Delta_{i,k_{i}}+1-\beta_{i,k_{i}}\end{array}$&
$\begin{array}{c}
\frac{\tau}{2}+\delta_{i}=:\frac{\tau}{2}+\Delta_{i,0}\\
\hD{\tau}{1}\\
\hD{\beta_{i,1}}{2}\\
\vdots\\
\hD{\beta_{i,k_{i}-1}}{k_{i}}\\
\mbox{reset:}\,\,1\mapsto0\end{array}$\tabularnewline
\end{tabular}\end{center}

\caption{Time evolution of oscillator $i$ in response to $k_{i}$ successively
incoming signals from its presynaptic oscillators $j_{n}$, $n\in\{1,\ldots,k_{i}\}$,
from which $i$ receives the $n^{\mathrm{th}}$ signal during this
period. The right column gives the phases $\phi_{i}(t)$ at times
$t$ given in the left column. The time evolution is shown for a part
of one period ranging from $\phi_{i}\approx\tau/2$ to reset, $1\reset0$,
such that $\phi_{i}=0$ in the last row. The first row gives the initial
condition $\phi_{i}(0)=\tau/2+\delta_{i}\,$. The following rows describe
the reception of the $k_{i}$ signals during this period whereby the
phases are mapped to $\beta_{i,n}$ after the $n^{\mathrm{th}}$ signal
has been received. The last row describes the reset at threshold such
that the respective time $T_{i}^{(0)}=\tau/2-\Delta_{i,k_{i}}+1-\beta_{i,k_{i}}$
gives the time to threshold of oscillator $i$.}

\label{Tab:complexmap}
\end{table}
 The time to threshold of oscillator $i$, which is given in the lower
left entry of the scheme, \begin{equation}
T_{i}^{(0)}:=\frac{\tau}{2}-\Delta_{i,k_{i}}+1-\beta_{i,k_{i}}\label{eq:time_to_spike}\end{equation}
 is about $\phi_{0}(0)=\tau/2$ smaller than the period $T$. Hence
the period-$T$ map of the perturbation can be expressed as \begin{equation}
\delta_{i}(T)=T-T_{i}^{(0)}-\frac{\tau}{2}=\beta_{i,k_{i}}-\alpha+\Delta_{i,k_{i}}\label{eq:delta_i}\end{equation}
where $\alpha$ is given by Eq.\ (\ref{eq:alpha}).

\section{Multiple first order operators}

In order to perform a local stability analysis, we consider the first
order approximations of the maps derived in the previous section.
Expanding $\beta_{i,k_{i}}$ for small $D_{i,n}\ll1$ one can proof
by induction \cite{Timme:2006:NC} that\begin{equation}
\beta_{i,k_{i}}\doteq\alpha+\sum_{n=1}^{k_{i}}p_{i,n-1}D_{i,n}\label{eq:beta_i}\end{equation}
 where \begin{equation}
p_{i,n}:=\frac{U'(U^{-1}(U(\tau)+\sum_{m=1}^{n}\eps_{ij_{m}}))}{U'(U^{-1}(U(\tau)+\eps))}\label{eq:fractions}\end{equation}
 for $n\in\{0,\,1,\,\ldots,k_{i}\}$ encodes the effect of an individual
incoming signal of strength $\eps_{ij_{n}}$. The statement $x\doteq y$
means that $x=y+\sum_{i,n}\mathcal{O}(D_{i,n}^{2})$ as all $D_{i,n}\rightarrow0$.
Substituting the first order approximation Eq.\ (\ref{eq:beta_i})
into Eq.\ (\ref{eq:delta_i}) using Eq.\ (\ref{eq:D_i,n}) leads
to\begin{eqnarray}
\delta_{i}(T)\doteq\sum_{n=1}^{k_{i}}p_{i,n-1}(\Delta_{i,n-1}-\Delta_{i,n})+\Delta_{i,k_{i}}\label{eq:delta_i(T)_for_sum}\end{eqnarray}
such that after rewriting \begin{equation}
\delta_{i}(T)\doteq p_{i,0}\Delta_{i,0}+\sum_{n=1}^{k_{i}}(p_{i,n}-p_{i,n-1})\Delta_{i,n}\label{eq:delta_i(T)_for_sum2}\end{equation}
to first order in all $\Delta_{i,n}$. Since $\Delta_{i,n}=\delta_{j_{n}(i)}$
for $n\in\{1,\ldots,k_{i}\}$ and $\Delta_{i,0}=\delta_{i}$ according
to Eqs.\ (\ref{eq:Deltaindeltajn}) and (\ref{eq:Deltai0}), this
results in a first order map\begin{equation}
\boldsymbol{\delta}(T)\doteq A\boldsymbol{\delta}\label{eq:matrixequation}\end{equation}
 where the elements of the matrix $A$ are given by \begin{equation}
A_{ij}=\left\{ \begin{array}{ll}
p_{i,n}-p_{i,n-1} & \mbox{if}\, j=j_{n}\in\Pre(i)\\
p_{i,0} & \mbox{if}\, j=i\\
0 & \mbox{if}\, j\notin\Pre(i)\cup\{ i\}.\end{array}\right.\label{eq:matrixelements}\end{equation}
 As for the nonlinear stroboscopic maps (\ref{eq:delta_i}), because
$j_{n}$ in Eq.\ (\ref{eq:matrixelements}) identifies the $n^{\mathrm{th}}$
pulse received during this period by oscillator $i$, the first order
operator depends on the rank order of the perturbations, $A=A(\rank(\boldsymbol{\delta}))$.
 The variables $p_{i,n}$ encode phase jumps evoked by all pulses
up to the $n\textrm{th}$ one received. Since the matrix elements
(\ref{eq:matrixelements}) are differences of these $p_{i,n\,},$
matrix elements $A_{i,j}$ and $A_{i,j'}$ with $j\neq j'$ have in
general different values depending on the order of incoming signals. 

This multi-operator problem is induced by the structure of the network
together with the pulsed interactions. For networks with homogeneous,
global coupling different matrices $A$ can be identified by an appropriate
permutation of the oscillator indices. In general, however, this is
impossible. Thus even for a network of given number of neuronal oscillators
at given connection strengths and given delay and interaction function,
the stability of the synchronous state is described by many different
operators that depend on the rank order of the perturbation.

\section{Alternative method \newline to determine stability }

In most stability problems for periodic orbits in dynamical systems
theory, finding the eigenvalues of an appropriate stroboscopic map
is sufficient for determining the stability of the orbit. Typically,
one eigenvalue equals one and corresponds to perturbation along the
periodic orbit trajectory such that there is no restoring force. If
all other eigenvalues are smaller than one in absolute value, the
periodic orbit is asymptotically stable and all sufficiently close
initial states converge to it. 

On the contrary, the multi-operator property of the stability problem
considered here implies that standard eigenvalue analysis fails. However,
we found other methods to determine the stability of the synchronous
periodic state. We present the results briefly in the following. 

To show plain (non-asymptotic) linear stability, observe that the
row-sums of the stability matrices are normalized, \begin{equation}
\sum_{j=1}^{N}A_{ij}=1\label{eq:sum_A_ij_1}\end{equation}
reflecting the invariance of the periodic orbit with respect to perturbations
along it. Given that the coupling strengths are purely inhibitory,
$\eps_{ij}\leq0$ , one can show that the $p_{i,n}$ (Eq.~(\ref{eq:fractions}))
are positive and bounded above by one,\begin{equation}
0<p_{i,n}\leq1,\end{equation}
and that they increase with $n$,\begin{equation}
p_{i,n-1}<p_{i,n}\,.\end{equation}
 Hence the nonzero off-diagonal elements are positive, $A_{ij_{n}}=p_{i,n}-p_{i,n-1}>0$
such that\begin{equation}
A_{ij}\geq0\label{eq:A_ij_inh_positive}\end{equation}
for all $i,j\in\{1,\ldots,N\}$. Moreover the diagonal elements \begin{equation}
A_{ii}=p_{i,0}=\frac{U'(\tau)}{U'(U^{-1}(U(\tau)+\eps))}=:A_{0}\label{eq:A_ii}\end{equation}
are identical for all $i$ and satisfy \begin{equation}
0<A_{0}<1\label{eq:A0inh}\end{equation}
 because $U$ is monotonically increasing, $U'(\phi)>0$, and concave
down, $U"(\phi)<0$, for all $\phi$. It is important to note that
$A$ has the properties Eqs.\ (\ref{eq:sum_A_ij_1})--(\ref{eq:A0inh})
independent of the parameters, the network connectivity, and the specific
perturbation considered. With these observations, it is straightforward
to show that the synchronous state is stable in the sense that small
perturbations cannot grow: To first order, a given perturbation $\boldsymbol{\delta}=\boldsymbol{\delta}(0)$
satisfies

\begin{eqnarray}
\left\Vert \boldsymbol{\delta}(T)\right\Vert  & := & \max_{i}|\delta_{i}(T)|\\
 & = & \max_{i}\left|\sum_{j=1}^{N}A_{ij}\delta_{j}\right|\\
 & \leq & \max_{i}\sum_{j}|A_{ij}||\delta_{j}|\\
 & \leq & \max_{i}\sum_{j}|A_{ij}|\max_{k}|\delta_{k}|\\
 & = & \max_{i}\sum_{j}A_{ij}\max_{k}|\delta_{k}|\\
 & = & \max_{k}|\delta_{k}|\\
 & = & \left\Vert \boldsymbol{\delta}\right\Vert \end{eqnarray}
where we use the vector norm \begin{equation}
\left\Vert \boldsymbol{\delta}\right\Vert :=\max_{i}|\delta_{i}|\,.\end{equation}
 Thus the length of a perturbation vector does not increase during
one period implying that it does not increase for an arbitrary long
time. Using methods from graph theory, one can show \cite{Timme:2006:NC}
that for strongly connected networks (in which every oscillator can
be reached from every other by following a directed path on the network)
the synchronous state is asymptotically stable such that from sufficiently
close initial conditions the spiking activity will become exactly
synchronous. The results on asymptotic stability use the recurrence
properties of strongly connected networks and rely on the fact that
every oscillator can communicate with every other at least indirectly.
Thus the results for plain and asymptotic stability are independent
of the specific choice of parameters, $\eps_{ij}\leq0$, $\tau\in(0,1)$,
the potential function $U(\phi)$, and the rank order of the perturbation.
They are derived without using the eigenvalues or eigenvectors of
a given stability matrix and solve the stability problem exactly.
In summary this means that \emph{any} network of the type described
above, with normalized inhibitory coupling Eq.\ (\ref{eq:epsnorm})
exhibits a synchronous state that is at least marginally stable; it
is moreover asymptotically stable if the network is strongly connected.

\begin{figure}[htp!]
\begin{center}\includegraphics[%
  width=6cm]{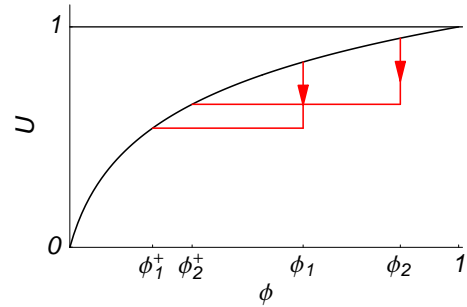}\end{center}

\caption{{\small Intuitive synchronization mechanism: Inhibition synchronizes
due to the concavity of $U$. Simultaneously received inhibitory input
decreases phase differences between the receiving oscillators,} {\small $\left|\phi_{2}(t^{+})-\phi_{1}(t^{+})\right|>\left|\phi_{2}(t)-\phi_{1}(t)\right|$
. \label{Fig:inhsyn}}}
\end{figure}

A simple intuitive argument why networks of inhibitorily coupled neural
oscillators synchronize can be obtained from the response dynamics
of individual units, Fig.~\ref{Fig:inhsyn}. If two (or more) neurons
simultaneously receive inhibitory input of the same size, their potential
is decreased by the same amount such that their potential difference
stays unchanged. Due to the negative curvature of the rise function
that mediates the negative input, this, however, leads to a decrease
of their phase differences, which encode the future spike times. %
 This intuitive explanation holds for simple situations like globally
coupled systems with homogeneous coupling strengths. However, the
synchronization dynamics is more complicated if the inputs are not
of equal size or only one input exists for some unit, as e.g., in
a ring of neurons.

In the case of integrate and fire rise functions $U=U_{\textrm{IF}}$
one can derive \cite{Timme:2006:NC} stability results based on the
eigensystem because all stability operators become degenerate; see
below for details of the degeneracy for networks of integrate and
fire neurons. The \Gershgorin disk theorem then bounds all eigenvalues
in a disk of radius $r_{\textrm{G}}=1-A_{0}$ centered at $A_{0}$,
touching the unit circle from the inside at $z=1$. It ensures that
the eigenvalue largest in magnitude is $\lambda_{1}=1$, with corresponding
eigenvector $\boldsymbol{{v}}_{1}\propto(1,1,\ldots,1)^{\mathsf{{T}}}$.
For strongly connected networks, the Perron Frobenius theorem implies
that this eigenvalue is unique, i.\ e.\  all other eigenvalues are
smaller than one in absolute value. Thus all perturbations that contain
components other than $\boldsymbol{{v}}_{1}$ will decay towards a
uniform perturbation. Such an analysis confirms that networks of arbitrary
connectivity are at least marginally stable and strongly connected
networks exhibit asymptotically stable synchronous states, as shown
above by alternative methods. 

If networks consist of several strongly connected components, the
analysis is much more involved and structural identification of strongly
connected components and the wiring among them is required. Such networks
display a novel kind of non-synchronous activity that is controlled
by the coarse and fine scale structure of the network, cf. \cite{Timme:Topid:2006}.
This state seems to be universal among networks of coupled oscillators
exhibiting a synchronization mechanism.

\section{Enforcing degeneracy:\protect \\
The phoenix integrate-and-fire }

From the general class of concave increasing rise functions, we now
derive a subclass of rise functions in which all multiple operators
degenerate to a single stability matrix if the coupling strength are
suitably chosen. Interestingly, it turns out that the class of standard
leaky integrate-and-fire oscillators provides potential functions
consistent with this condition. 

A general potential function $U$ that is monotonically increasing,
$U'(\phi)>0$, and concave (down), $U''(\phi)<0$, yielded stability
operators $A$ in the first order map (\ref{eq:matrixequation})
that are defined by their respective matrix elements (\ref{eq:matrixelements})
in terms of differences of the $p_{i,n}$ (Eq.\ \ref{eq:fractions})
that in turn describe the effect of the $n\textrm{th}$ signal received
by oscillator $i$ within the period considered. Thus, the actual
stability operator to be used for a specific perturbation depends
on the rank order of the incoming signals given this perturbation.
Can the multiple linear operators be made degenerate? If so, the eigensystem
of the resulting matrix completely described the asymptotic synchronization
dynamics.

Consider a network for which the coupling strengths of all presynaptic
oscillators $j\in\Pre(i)$ are identical,\begin{equation}
\eps_{ij}=\frac{\eps}{k_{i}}\textrm{ }\label{eq:epsijepski}\end{equation}
for each oscillator $i$. For such a network, two matrix elements
are interchanged at the boundary of the domains of definition of an
individual operator. For instance, assume that an oscillator $i$
has exactly two presynaptic oscillators $j$ and $j'$. If a perturbation
is changed such that $\delta_{j}>\delta_{j'}$ is turned into $\delta_{j}<\delta_{j'}$,
the operator $A$ will change from $A=A^{(1)}$ to $A=A^{(2)}$ where
the non-zero off-diagonal elements of row $i$ read \begin{eqnarray}
A_{ij}^{(1)}=A_{ij_{1}}=p_{i,1}-p_{i,0} & \textrm{;}\, & A_{ij'}^{(1)}=A_{ij_{2}}=p_{i,2}-p_{i,1}\,,\\
A_{ij}^{(2)}=A_{ij_{2}}=p_{i,2}-p_{i,1} & \textrm{;\,} & A_{ij'}^{(2)}=A_{ij_{1}}=p_{i,1}-p_{i,0}\,,\end{eqnarray}
respectively. As above, $j_{1}$ labels the oscillator presynaptic
to $i$ that has sent the first signal to $i$ during the period considered,
and $j_{2}$ labels the presynaptic oscillator that has sent the second
one such that\begin{eqnarray}
j_{1}=j\textrm{ and }j_{2}=j' & \quad\Leftrightarrow\quad & \delta_{j}>\delta_{j'}\,,\\
j_{1}=j'\textrm{ and }j_{2}=j & \quad\Leftrightarrow\quad & \delta_{j'}>\delta_{j}\,.\end{eqnarray}
 Degeneracy of these two, in general distinct, cases requires that\begin{equation}
A_{ij}^{(k)}\overset{!}{=}A_{ij}^{(l)}\label{eq:AijkAijl}\end{equation}
 for $k,l\in\{1,2\}$ or, equivalently,\begin{equation}
p_{i,2}-p_{i,1}\overset{!}{=}p_{i,1}-p_{i,0}\,.\end{equation}

If every oscillator $i\in\{1,\ldots,N\}$ in the network has $k_{i}$
presynaptic oscillators, the degeneracy condition is easily generalized
to Eq.\ (\ref{eq:AijkAijl}) with $k$ and $l$ running over all
different stability matrices that occur for all possible differently
ordered perturbations. Expressed in terms of the $p_{i,n}$, which
describe the effect of individual incoming pulses, we obtain \begin{equation}
p_{i,n}-p_{i,n-1}\overset{!}{=}p_{i,m}-p_{i,m-1}\label{eq:pinpim}\end{equation}
 for all $i\in\{1,\ldots,N\}$ and all $n,m\in\{1,\ldots,k_{i}\}$.

If we define \begin{equation}
q(x_{i,n}):=p_{i,n}=\frac{U'(U^{-1}(U(\tau)+x_{i,n}))}{U'(U^{-1}(U(\tau)+\eps))}\label{eq:q(x)}\end{equation}
where \begin{equation}
x_{i,n}=\sum_{m=1}^{n}\eps_{ij_{m}}=\frac{n\eps}{k_{i}}\label{eq:x}\end{equation}
for $n\leq k_{i}$, the requirement (\ref{eq:pinpim}) is satisfied
if\begin{equation}
q'(x)=\textrm{const}\label{eq:qprimeconstant}\end{equation}
 in the relevant interval $x\in[\eps,0]$. Note that $\eps<0$ because
we consider inhibitory coupling. The first derivative of $q(x)$ satisfies\begin{equation}
q'(x)\propto\frac{U''(U^{-1}(U(\tau)+x))}{U'(U^{-1}(U(\tau)+x))}=:\frac{U''(h(x))}{U'(h(x))}\label{eq:q_prime_const}\end{equation}
where $h(x)=U^{-1}(U(\tau)+x)$ is an invertible function of $x$.
Together with Eq.\ (\ref{eq:qprimeconstant}) this leads to a differential
equation \begin{equation}
U''=cU'\end{equation}
where $c\in\mathbb{R}$ is a constant. The solution $U(\phi)=a+be^{c\phi}$
with constants $a,b,c\in\mathbb{R}$ together with the normalization
$U(0)=0$, $U(1)=1$, and the monotonicity and concavity requirements,
$U'(\phi)>0$ and $U"(\phi)<0$, yield the one-parameter family of
solutions in integrate-and-fire form\begin{equation}
U(\phi)=U_{\textrm{IF}}(\phi)=I(1-e^{-\phi T_{\textrm{IF}}})\end{equation}
where $I>1$ and $T_{\textrm{IF}}=\ln(I/(I-1))>0$. This leads to\begin{equation}
U_{\textrm{IF}}'(\phi)=IT_{\textrm{IF}}e^{-\phi T_{\textrm{IF}}},\end{equation}
\begin{equation}
U_{\textrm{IF}}^{-1}(y)=\frac{1}{T_{\textrm{IF}}}\ln\left(1-\frac{y}{I}\right)^{-1},\end{equation}
and \begin{equation}
U_{\textrm{IF}}^{-1}(U_{\textrm{IF}}(\phi)+\eps)=\frac{1}{T_{\textrm{IF}}}\ln\left(e^{-\phi T_{\textrm{IF}}}-\frac{\eps}{I}\right)^{-1}\end{equation}
such that\begin{equation}
U_{\textrm{IF}}'(U_{\textrm{IF}}^{-1}(U(\phi)+\eps))=T_{\textrm{IF}}\left(Ie^{-\phi T_{\textrm{IF}}}-\eps\right)\end{equation}
and\begin{eqnarray}
p_{i,n} & = & \frac{U_{\textrm{IF}}'(U_{\textrm{IF}}^{-1}(U_{\textrm{IF}}(\tau)+\sum_{m=1}^{n}\eps_{ij_{m}}))}{U_{\textrm{IF}}'(U_{\textrm{IF}}^{-1}(U_{\textrm{IF}}(\tau)+\eps))}\\
 & = & \frac{Ie^{-\tau T_{\textrm{IF}}}-\sum_{m=1}^{n}\eps_{ij_{m}}}{Ie^{-\tau T_{\textrm{IF}}}-\eps}\,.\end{eqnarray}
 Thus, by construction, if we substitute $\eps_{ij_{n}}=\eps/k_{i}$
all non-zero off-diagonal elements \begin{equation}
A_{ij_{n}}=p_{i,n}-p_{i,n-1}=\frac{1}{Ie^{-\tau T_{\textrm{IF}}}-\eps}\frac{\eps}{k_{i}}\label{eq:AijIF}\end{equation}
 in one row $i$ of the stability matrix are identical,\begin{equation}
A_{ij_{n}}=A_{ij_{m}},\label{eq:AijnAijm}\end{equation}
 for all $n,m\in\{1,\ldots,k_{i}\}$. 

One should note that, given the coupling strengths satisfy Eq.\ (\ref{eq:epsijepski}),
the condition (\ref{eq:qprimeconstant}) is sufficient but not necessary
for degeneracy of all operators. At given parameters and a given network
connectivity, one can construct potential functions that fulfill condition
(\ref{eq:qprimeconstant}) only on (local) average such that the requirement
for identical (non-zero) off-diagonal matrix elements in each row
(\ref{eq:pinpim}) is still satisfied. If we do not a priori fix the
parameters and the network structure, however, the potential function
$U_{\textrm{IF}}$ uniquely leads to operator degeneracy within the
class of concave down, increasing functions.

This degeneracy has important consequences: Whereas for the multi-operator
problem the dynamics in the vicinity of the synchronous state is determined
by an (unknown but deterministic) sequence of different linear operators,
the dynamics in case of degeneracy is determined by the eigenvectors
and eigenvalues of a single matrix $A$. In particular, the second
largest eigenvalue\begin{equation}
\lambda_{\textrm{m}}:=\max\{|\lambda_{i}|\,:\,\,|\lambda_{i}|<1\}\label{eq:lambdam}\end{equation}
of this matrix $A$ determines the asymptotic speed of convergence
towards the synchronous state,\begin{equation}
|\boldsymbol{\delta}((n+l)T)|\sim\lambda_{\textrm{m}}^{n}|\boldsymbol{\delta}(lT)|\end{equation}
 for $n,l\gg1$.

Interestingly, the derivation of a condition for degeneracy led to
the standard leaky integrate-and-fire model as a subclass of models
that imply degeneracy for suitably chosen coupling strengths. Starting
from this degenerate case of operators now enables us to develop a
characterization of the synchronization dynamics in terms of eigenvalues
of that operator.

\section{Location of eigenvalues \protect \\
in large random networks}

\label{sec:locationnumerics}

How fast do random networks synchronize? The characteristic asymptotic
time of synchronization, $\tau_{\textrm{syn}}=-1/\ln(\lambda_{\textrm{m}})$,
see Eq.\ (\ref{eq:tausyn}) below, is given in terms of the second
largest eigenvalue $\lambda_{\textrm{m}}$ that we determine from
the distribution of eigenvalues in the following sections. In this
section, we present examples for the distribution of eigenvalues of
stability matrices describing the asymptotic dynamics of large asymmetric
random networks of integrate-and-fire oscillators in the vicinity
of the synchronous state. From the details of the analysis described
above, we know that all eigenvalues must be located in a \Gershgorin 
disk $K$ in the complex plane that is centered at $A_{0}<1$ (Eq.\
\equa{A0inh}) and has radius $1-A_{0}$ such that it contacts the
unit circle at $z=1$ from the inside. In the following, we consider
neural oscillators that interact inhibitorily on two classes of random
networks (defined in subsections \ref{subsec:fixedindegree} and \ref{subsec:fixedprob}).
The potential functions of the oscillators are of the integrate-and-fire
form $U(\phi)=U_{\textrm{IF}}(\phi)=I(1-e^{-\phi T_{\textrm{IF}}})$,
where $T_{\textrm{IF}}=\ln(I/(I-1))$. The non-zero coupling strength
are chosen according to $\eps_{ij}=\eps/k_{i\,}$. We consider only
sparsely connected networks which lead to sparse stability matrices
where we term a matrix {}``sparse'' if at least a positive fraction
of its entries is zero in the limit of large $N$.

\subsection{Networks with constant in-degree}

\label{subsec:fixedindegree}

\begin{figure*}[htp!]
\begin{center}\includegraphics[%
  width=16cm]{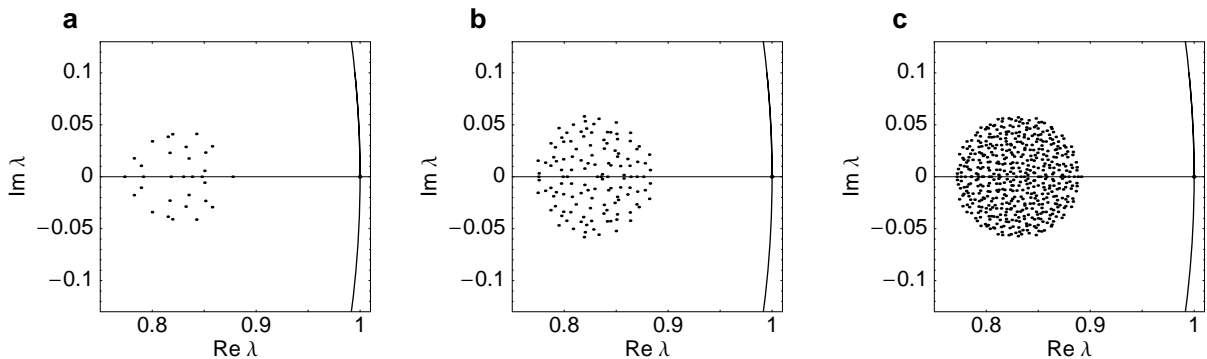}\end{center}

\caption{Distribution of eigenvalues $\lambda_{i}$ in the complex plane for
networks of fixed in-degree $k=8$ and different sizes (a) $N=32$,
(b) $N=128$, (c) $N=512$. For large networks, the non-trivial eigenvalues
seem to be distributed uniformly on a disk in the complex plane. The
arc through the trivial eigenvalue (dot at $z=\lambda_{1}=1$) is
a sector of the unit circle. Parameters of integrate-and-fire oscillators
are $I=1.1$, $\eps=-0.2$, $\tau=0.05$. \label{Fig:EWSfreek8N}}
\end{figure*}

The first class of networks is given by random networks in which all
oscillators $i$ have the same number $k_{i}=k$ of presynaptic oscillators
which are independently drawn from the set of all other oscillators
with uniform probability. When increasing the network size $N$, the
number of connections $k$ per oscillator is kept fixed. We numerically
determined the eigenvalues of different stability matrices changing
the network size $N\in\{2^{6},\ldots,2^{14}\}$, the in-degree $k\in\{2,\ldots2^{8}\}$,
and the dynamical parameters $\eps$, $\tau$, and $I$ such that
$A_{0}\in[0.6,0.9]$. In general, we find that, for sufficiently large
$N$ and sufficiently large $k$, the non-trivial eigenvalues resemble
a disk in the complex plane that is centered at about $A_{0}$ but
has a radius $r$ that is smaller than the upper bound given by the
\Gershgorin  theorem\begin{equation}
r<1-A_{0}.\label{eq:rsmaller1A0}\end{equation}
Note that, due to the invariance of the periodic orbit with respect
to globally constant phase shifts, there is always a trivial eigenvalue
$\lambda_{1}=1$. As an example, the eigenvalue distributions in the
complex plane are displayed in Fig.~\ref{Fig:EWSfreek8N} for specific
parameters and differently sized networks.

\subsection{Networks with constant connection probability}

\label{subsec:fixedprob}

\begin{figure*}[htp!]
\begin{center}\includegraphics[%
  width=16cm]{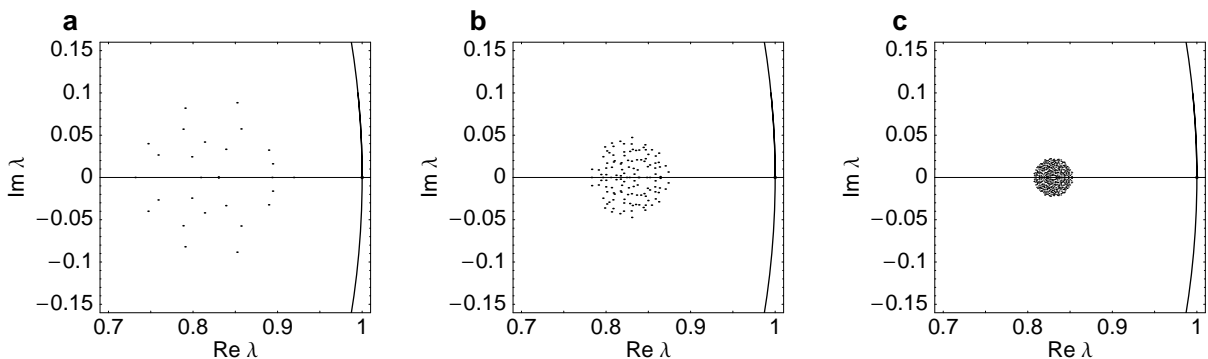}\end{center}

\caption{Distribution of eigenvalues $\lambda_{i}$ in the complex plane for
networks of fixed connection probability $p=0.1$ and different sizes
(a) $N=32$, (b) $N=128$, (c) $N=512$. For large networks, the non-trivial
eigenvalues seem to be distributed uniformly on a disk in the complex
plane, the radius of which shrinks with increasing network size. The
arc through the trivial eigenvalue (dot at $z=\lambda_{1}=1$) is
a sector of the unit circle. Parameters of integrate-and-fire oscillators
are $I=1.1$, $\eps=-0.2$, $\tau=0.05$. \label{Fig:EWSfreep0.1N}}
\end{figure*}

The second class of networks is given by random networks for which
every connection between any oscillator $i$ and any other oscillator
$j\neq i$ is present with given probability $p$. When increasing
$N$, this probability is kept fixed such that the number of connections
per oscillator is proportional to $N$. As for the other class of
random networks, we find numerically that the distribution of non-trivial
eigenvalues resemble disks in the complex plane that are smaller than
the \Gershgorin  disk (\ref{eq:rsmaller1A0}) but centered at about
the same point $A_{0}$. We numerically determined the distribution
of eigenvalues for $N\in\{2^{8},\ldots,2^{14}\}$, $p\in[0.01,0.2]$,
and the dynamical parameters $\eps$, $\tau$, and $I$ such that
$A_{0}\in[0.6,0.9]$. Figure \ref{Fig:EWSfreep0.1N} displays examples
of eigenvalue distributions for differently sized networks at the
otherwise identical parameters.

\section{Predictions from Random Matrix Theory}

The results of the previous section indicate that the eigenvalues
of stability matrices for large asymmetric random networks of integrate-and-fire
oscillators are located in disks in the complex plane if the network
size $N$ is sufficiently large. If this could be demonstrated independent
of specific parameters, it would be guaranteed that all non-trivial
eigenvalues are separated from the unit circle. Thus the main condition
required for the robustness of the stable synchronous state under
a structural perturbation to the dynamics of the system would be satisfied.
Moreover, the asymptotic synchronization time can be predicted analytically
from these results.

How can we predict the location of the eigenvalues? Since we are considering
random networks, a natural starting point is the theory of random
matrices. Random Matrix Theory has been investigated intensively since
the early 1950s \cite{Wigner:1951:790} (see also \cite{Porter:1965,Mehta:1991})
and turned out to be a valuable tool for both qualitative and quantitative
description of spectral properties of complex systems. For instance,
it describes level correlations in nuclear physics \cite{Guhr:1998:189}
as well as quantum mechanical aspects of chaos \cite{Bohigas:1984,Haake:2001}.
In applications of Random Matrix Theory to physical problems, it is
generally assumed that the details of the physical system are less
important for many statistical properties of interest. Often it turns
out that important statistical properties such as the distribution
of spacings of energy levels in quantum systems are well described
by the respective properties of random matrices that respect the same
symmetries as the physical system. Both theoretical investigations
and applications of Random Matrix Theory have focused on symmetric
matrices. Asymmetric matrices are less well understood and found only
limited applicability. In the following, we will evaluate the applicability
of Random Matrix Theory for estimating distributions of eigenvalues
of asymmetric stability matrices.

\subsection{Ensembles of symmetric and asymmetric random matrices}

For the case of real \textit{symmetric} random $N\times N$-matrices
$J=J^{\mathsf{T}}$ with independent, identically distributed components
$J_{ij}=J_{ji}\,$, it is believed \cite{Mirlin:1991:2273,Fyodorov:1991:2049}
that there are exactly two universality classes. Every ensemble of
matrices within one of these universality classes exhibits the same
distribution of eigenvalues in the limit of large matrices, $N\rightarrow\infty$,
but the eigenvalue distributions are in general different for the
two classes. Both universality classes are characterized by specific
ensembles of matrices the elements of which are distributed according
to a simple probability distribution. The class of sparse matrices
is represented by the probability distribution\begin{equation}
p_{\textrm{sparse}}(J_{ij})=\frac{k}{N}\delta\left(J_{ij}-\frac{1}{k}\right)+\left(1-\frac{k}{N}\right)\delta(J_{ij})\label{eq:prob_symmetric}\end{equation}
 where $k$ is the (finite) average number of entries in any row $i$
and $\delta(\cdot)$ is the Dirac delta distribution. The class of
Gaussian random matrices is exemplified by a Gaussian distribution
of matrix elements \begin{equation}
p_{\textrm{Gauss}}(J_{ij})=N^{\frac{1}{2}}(2\pi s^{2})^{-\frac{1}{2}}\exp\left(-\frac{NJ_{ij}^{2}}{2s^{2}}\right).\label{eq:prob_Gauss}\end{equation}
To obtain symmetric matrices, one chooses $J_{ij}=J_{ji}$ and $J_{ii}=0$
for both ensembles. Thus the arithmetic mean of the eigenvalues is
zero,\begin{equation}
\left[\lambda_{i}\right]_{i}:=\frac{1}{N}\sum_{i=1}^{N}\lambda_{i}=\frac{1}{N}\sum_{i=1}^{N}J_{ii}=0\label{eq:meanlambda0}\end{equation}
 and the ensemble variance of the matrix elements scale like \begin{equation}
\sigma^{2}=\left\langle J_{ij}^{2}\right\rangle \doteq\frac{r^{2}}{N}\label{eq:Gaussian_variance}\end{equation}
for $N\gg1$. For the Gaussian symmetric ensemble, it is known \cite{Wigner:1951:790,Mehta:1991}
that the distribution of eigenvalues $\rho_{\textrm{Gauss}}^{\textrm{s}}(\lambda)$
in the limit $N\rightarrow\infty$ is given by Wigner's semicircle
law \begin{equation}
\rho_{\textrm{Gauss}}^{\textrm{s}}(\lambda)=\left\{ \begin{array}{cc}
\frac{1}{2\pi r^{2}}(4r^{2}-\lambda^{2})^{\frac{1}{2}} & \textrm{if }|\lambda|\leq2r\\
0 & \textrm{otherwise}.\end{array}\right.\end{equation}
The ensemble of sparse matrices \cite{Bray:1988:11461,Rodgers:1988:3557,Fyodorov:1991:2049,Mirlin:1991:2273}
exhibits a different eigenvalue distribution $\rho_{\textrm{sparse}}^{\textrm{s}}(\lambda)$
that depends on the finite number $k$ of nonzero entries per row
and approaches the distribution $\rho_{\textrm{Gauss}}^{\textrm{s}}(\lambda)$
in the limit of large $k$ such that\begin{equation}
\lim_{k\rightarrow\infty}\rho_{\textrm{sparse}}^{\textrm{s}}(\lambda)=\rho_{\textrm{Gauss}}^{\textrm{s}}(\lambda).\label{eq:sparseGauss}\end{equation}
 It is important to note that in the limit of large $N$ the distributions
$\rho_{\textrm{sparse}}^{\textrm{s}}$ and $\rho_{\textrm{Gauss}}^{\textrm{s}}$
eigenvalues depend only on the one parameter $r$, that is derived
from the variance of the matrix elements (\ref{eq:Gaussian_variance}).

For real, \textit{asymmetric} matrices (independent $J_{ij}$ and
$J_{ji}$), there are no analytical results for the case of sparse
matrices but only for the case of Gaussian random matrices. The Gaussian
asymmetric ensemble (e.g.\  Eq.\ (\ref{eq:prob_Gauss}) with independent
$J_{ij}$ and $J_{ji}$) yields the distribution of complex eigenvalues
in a disk in the complex plane \cite{Girko:1985:694,Sommers:1988:1895}\begin{equation}
\rho_{\textrm{Gauss}}^{\textrm{a}}(\lambda)=\left\{ \begin{array}{ll}
(\pi r^{2})^{-1} & \textrm{if }|\lambda|\leq r\\
0 & \textrm{otherwise}\end{array}\right.\label{eq:uniform_EW_distribution}\end{equation}
where $r$ from Eq.\ (\ref{eq:Gaussian_variance}) is the radius
of the disk that is centered at zero. Like in the case of symmetric
matrices, this distribution also depends only on one parameter $r$,
that is derived from the variance of the matrix elements.

\subsection{Stability matrices and the Gaussian asymmetric ensemble}

In the numerical studies of stability matrices for random networks
(Sec.~\ref{sec:locationnumerics}), we observed that all non-trivial
eigenvalues of sparse stability matrices $A$ are located on or near
a disk in the complex plane (Figures \ref{Fig:EWSfreek8N} and \ref{Fig:EWSfreep0.1N}).
Since this is also predicted by the theory of asymmetric Gaussian
random matrices, let us compare these predictions to numerical results.
If the distribution of eigenvalues of sparse asymmetric random matrices
$\rho_{\textrm{sparse}}^{\textrm{a}}$ for $k\gg1$ is approximately
equal to the distribution of Gaussian asymmetric matrices, $\rho_{\textrm{sparse}}^{\textrm{a}}(\lambda)\approx\rho_{\textrm{Gauss}}^{\textrm{a}}(\lambda)$,
in analogy to the case of symmetric matrices (\ref{eq:sparseGauss}),
and Random Matrix Theory is applicable to the stability matrices at
all, we can obtain an analytical prediction for the radii of the disks
of eigenvalues. 

The elements of the original stability matrix $A$ have an average
\begin{equation}
\left[A_{ij}\right]=\frac{1}{N}\sum_{j=1}^{N}A_{ij}=\frac{1}{N}\end{equation}
and a second moment\begin{equation}
\left[A_{ij}^{2}\right]=\frac{1}{N}\sum_{j=1}^{N}A_{ij}^{2}=\frac{1}{N}\left(A_{0}^{2}+\sum_{\mysubstack{j=1}{j\neq i}}^{N}A_{ij}^{2}\right)\end{equation}
where the off-diagonal sum is bounded above and below by\begin{equation}
\frac{(1-A_{0})^{2}}{\max_{i}k_{i}}\leq\sum_{\mysubstack{j=1}{j\neq i}}^{N}A_{ij}^{2}\leq(1-A_{0})^{2}\end{equation}
due to the normalization (\ref{eq:sum_A_ij_1}). 

The variance $\sigma_{A}^{2}=\left[A_{ij}^{2}\right]-\left[A_{ij}\right]^{2}$
given by \begin{equation}
\sigma_{A}^{2}=\frac{A_{0}^{2}}{N}+\frac{\sum_{j\neq i}A_{ij}^{2}}{N}-\frac{1}{N^{2}}\end{equation}
is thus also bounded\begin{equation}
\frac{A_{0}^{2}}{N}+\frac{(1-A_{0})^{2}}{N(N-1)}-\frac{1}{N^{2}}\leq\sigma_{A}^{2}\leq\frac{A_{0}^{2}+(1-A_{0})^{2}}{N}-\frac{1}{N^{2}}\label{eq:variance_A_bounds}\end{equation}
because $\max_{i}k_{i}\leq N-1$. The eigenvalues of the original
matrix $A$ have the average value \begin{equation}
\left[\lambda_{i}\right]:=\frac{1}{N}\sum_{i=1}^{N}\lambda_{i}=\frac{1}{N}\sum_{i=1}^{N}A_{ii}=A_{0}.\end{equation}
To directly compare the ensemble of the stability matrices considered
here to random matrices with zero average eigenvalue, $\left\langle \lambda_{i}\right\rangle =0$,
and given variance (\ref{eq:Gaussian_variance}), we transform the
stability matrix $A$ to\begin{eqnarray}
A'_{ij} & = & A_{ij}-A_{0}\delta_{ij}\end{eqnarray}
for $i\in\{1,\ldots,N\}$. Here $\delta_{ij}$ denotes the Kronecker
delta, $\delta_{ij}=1$ if $i=j$ and $\delta_{ij}=0$ if $i\neq j$.
The transformation to $A'$ shifts all eigenvalues by $-A_{0}$ and
hence the average value of the eigenvalues to \begin{equation}
\left[\lambda'_{i}\right]=0\,.\end{equation}
In addition \begin{equation}
\left[A'_{ij}\right]=\left[A_{ij}\right]-\frac{A_{0}}{N}=\frac{(1-A_{0})}{N}\end{equation}
and \begin{equation}
\left[A'_{ij}²\right]=\left[A_{ij}²\right]-\frac{A_{0}^{2}}{N}\end{equation}
such that the variance is\begin{eqnarray}
\sigma_{A'}^{2} & = & \sigma_{A}^{2}-\frac{A_{0}^{2}}{N}+\frac{2A_{0}}{N^{2}}-\frac{A_{0}^{2}}{N^{2}}\\
 & = & \frac{1}{N}\left(\sum_{\mysubstack{j=1}{j\neq i}}^{N}A_{ij}^{2}-\frac{(1-A_{0})^{2}}{N}\right).\label{eq:varianceA'}\end{eqnarray}

The eigenvalue distribution of this ensemble of rescaled stability
matrices $A'$ for random networks may be compared to the Gaussian
asymmetric ensemble with zero average eigenvalue and variance $\sigma_{A'}^{2}\,$.
In such a comparison, the additional eigenvalue $\lambda_{1}=1$ of
$A$, is neglected. This should not matter for large networks ($N\gg1$).

It is important to note that we compare the location of eigenvalues
of a \textit{sparse} matrix with deterministic non-zero entries at
certain random positions with the eigenvalue distribution of the \textit{Gaussian}
ensemble, which consists of fully occupied matrices with purely random
entries. 

If we assume that the eigenvalue distributions for these two ensembles
of networks with fixed in-degree and networks with a fixed connection
probability are similar to those for random matrices, we obtain a
prediction \begin{equation}
r^{2}\approx N\sigma_{A'}^{2}\label{eq:radiusGaussianradiussparse}\end{equation}
for the radius of the disk of eigenvalues from Eq.~(\ref{eq:Gaussian_variance}).
For further investigations, we consider the two exemplary classes
of large random networks of integrate-and-fire oscillators discussed
in Sec.~\ref{sec:locationnumerics}. If we assume that the stability
matrix $A$ has \textit{exactly} $k$ non-zero off-diagonal elements
per row and identical coupling strength $\eps_{ij}=\eps/k$ between
the integrate-and-fire oscillators, the off-diagonal sum is exactly
equal to \begin{equation}
\sum_{n=1}^{k}A_{ij_{n}}^{2}=\frac{(1-A_{0})^{2}}{k}.\end{equation}
such that the variance of $A$ equals \begin{equation}
\sigma_{A}^{2}=\frac{A_{0}^{2}}{N}+\frac{(1-A_{0})^{2}}{Nk}-\frac{1}{N^{2}}\label{eq:varianceA}\end{equation}
and the variance of $A'$ is given by \begin{equation}
\sigma_{A'}^{2}=\frac{1}{N}(1-A_{0})^{2}\left(\frac{1}{k}-\frac{1}{N}\right).\end{equation}
If we now take the prediction from Random Matrix Theory $r_{\textrm{RMT}}$
for the radius $r$ of the disk of eigenvalues of the stability matrices,
we obtain

\begin{equation}
r_{\textrm{RMT}}=N^{\frac{1}{2}}\sigma_{A'}=(1-A_{0})\left(\frac{1}{k}-\frac{1}{N}\right)^{\frac{1}{2}}.\label{eq:rRMT}\end{equation}
In random networks where all oscillators have \textit{exactly} $k$
presynaptic oscillators, the approximation for the variance of $A$
(and thus of $A'$) is exact. If the random network is constructed
by choosing every connection independently with probability $p$,
the variance (\ref{eq:varianceA}) is only an approximation because
we replaced $\left[k_{i}^{-1}\right]_{i}$ by $k^{-1}$ which gives
the order of magnitude of the number of connections as a function
of $N$.

Substituting $A_{0}=1-\sum_{j,j\neq i}A_{ij}$ for integrate and fire
neurons (\ref{eq:AijIF}) into the radius prediction $r_{\textrm{RMT}}$,
Eq.~(\ref{eq:rRMT}), we obtain\begin{equation}
r_{\textrm{RMT}}^{\textrm{IF}}=\left(\frac{\eps}{Ie^{-\tau T_{\textrm{IF}}}-\eps}\right)\left(\frac{1}{k}-\frac{1}{N}\right)^{\frac{1}{2}}\label{eq:rRMTIF}\end{equation}
which explicitely contains all parameters of the system.

\section{Numerical tests of eigenvalue predictions }

We verified this scaling law for different parameters $A_{0}$ determined
by different $I$, $\eps$, and $\tau$ and found good agreement with
numerically determined eigenvalue distributions. We compared the theoretical
prediction (\ref{eq:rRMT}) to the numerical data for both ensembles
considered in Sec.~\ref{sec:locationnumerics}.

\subsection{Networks with constant in-degree}

\begin{figure*}[htp!]
\begin{center}\includegraphics[%
  width=16cm]{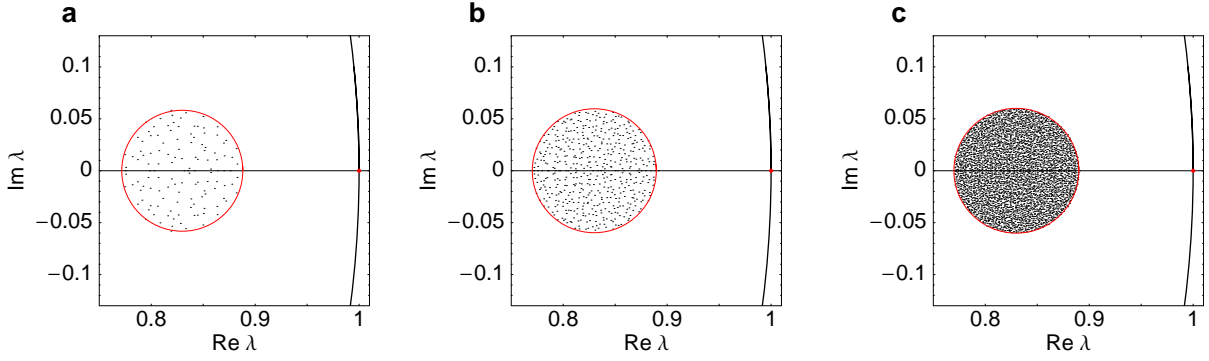}\end{center}

\caption{Distribution of eigenvalues in the complex plane for networks with
fixed in-degree $k=8$ for different network sizes (a) $N=128$, (b)
$N=512$, (c) $N=4096$. The disks are centered at $A_{0}$ and have
radius $r_{\textrm{RMT}}$, the prediction obtained from Random Matrix
Theory. The arc through the trivial eigenvalue $z=\lambda_{1}=1$
is a sector of the unit circle. Parameters of integrate-and-fire oscillators
are $I=1.1$, $\eps=-0.2$, $\tau=0.05$. \label{Fig:EWScirck8N}}
\end{figure*}

At a given network connectivity and given parameters, we obtained
all eigenvalues of the stability matrix $A$ for several network sizes
$N$ and in-degrees $k$. We find that the prediction obtained from
Random Matrix Theory well describes the numerically determined eigenvalues.
Examples of eigenvalue distributions for matrices at fixed $k$ and
three different $N$ are shown in Fig.~\ref{Fig:EWScirck8N}.

\begin{figure}[htp!]
\begin{center}\includegraphics[%
  width=85mm]{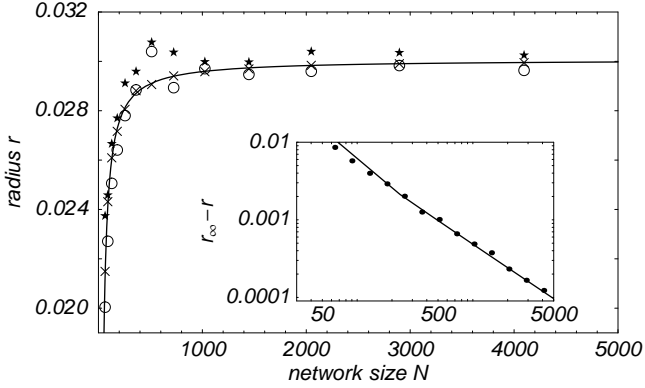}\end{center}

\caption{Scaling of the radius $r$ of the disk of non-trivial eigenvalues
with the network size $N$ at fixed in-degree $k=32$ ($I=1.1$, $\eps=-0.2$,
$\tau=0.05$). Main panel displays the radius $r$ as a function of
network size $N$. Symbols display $r_{\textrm{rad}}$ ($\star$),
$r_{\textrm{av}}$ ($\times$) and $r_{\textrm{Re}}$ ($\bigcirc$).
Inset displays $r_{\infty}-r$ as a function of $N$ on a doubly logarithmic
scale, where $r_{\infty}=(1-A_{0})k^{-1/2}$. Dots display numerical
data of $r_{\textrm{av}}$. In the main panel and the inset, lines
are the theoretical prediction $r_{\textrm{RMT}}=(1-A_{0})(1/k-1/N)^{1/2}$.\label{Fig:scalingk32N}}
\end{figure}

\begin{figure}[htp!]
\begin{center}\includegraphics[%
  width=85mm]{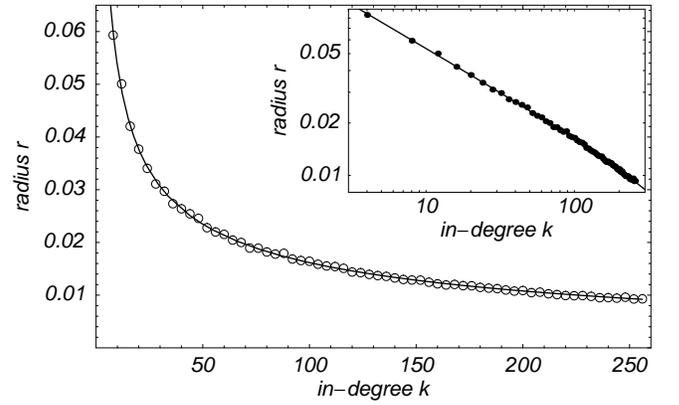}\end{center}

\caption{Scaling of the radius $r$ of the disk of non-trivial eigenvalues
with the in-degree $k$ for random networks of $N=1024$ oscillators
($I=1.1$, $\eps=-0.2$, $\tau=0.05$). Main panel displays the radius
$r$ as a function of in-degree $k$. Inset displays the same data
on a doubly logarithmic scale. Symbols display numerical results,
using the average estimator $r_{\textrm{av}}$, lines are the theoretical
prediction $r_{\textrm{RMT}}=(1-A_{0})(1/k-1/N)^{1/2}$.\label{Fig:scalingkN1024}}
\end{figure}

There are several ways to numerically estimate the radius of the disk
of eigenvalues. For illustration, we use three different estimators
here. The real part estimator \begin{equation}
r_{\textrm{Re}}:=\frac{1}{2}\left(\max_{i\neq1}\MyRe(\lambda_{i})-\min_{i\neq1}\MyRe(\lambda_{i})\right)\end{equation}
 estimates the radius from the maximum spread of eigenvalues parallel
to the real axis. Typically, $r_{\textrm{Re}}$ should give an estimate
that is too low compared to the radius obtained from the eigenvalues
of an ensemble of matrices because it measures the maximal spread
in one direction only. This is circumvented by the radial estimator\begin{equation}
r_{\textrm{rad}}:=\max_{i\neq1}|\lambda_{i}-(A_{0}-(1-A_{0})N^{-1})|\label{eq:rrad}\end{equation}
 that finds the maximum distance of any non-trivial eigenvalues from
the average of the non-trivial eigenvalues, $\left\langle \lambda_{i}\right\rangle _{i\neq1}=A_{0}-(1-A_{0})N^{-1}+\mathcal{O}(N^{-2})$.
This estimator should yield an approximation that may be too large
compared to the respective ensemble average. The average estimator\begin{equation}
r_{\textrm{av}}:=\frac{3}{2}\frac{1}{N-1}\sum_{i=2}^{N}|\lambda_{i}-(A_{0}-(1-A_{0})N^{-1})|\end{equation}
estimates the radius $r$ of a circle from the average distance $\left\langle d\right\rangle $
of eigenvalues from its center, because \begin{equation}
\left\langle d\right\rangle =\int_{0}^{2\pi}\int_{0}^{r}r'^{2}\rho(r')drd\varphi=\frac{2}{3}r\end{equation}
 if we assume a uniform $\rho(r')=1/(\pi r^{2})$ for $r'<r$ and
$\rho(r')=0$ otherwise (\ref{eq:uniform_EW_distribution}). This
estimate has the advantage, that it contains information from all
eigenvalues in contradistinction to the two other estimators. Its
disadvantage is that one has to assume a priori a uniform distribution
of non-trivial eigenvalues. As displayed in Fig.~\ref{Fig:scalingk32N},
all three estimators converge towards the radius predicted by the
random matrix model for large $N$ and given in-degree $k$.%
 Varying the in-degree $k$ at fixed $N$ also yields excellent agreement
between the numerical data and the theoretical predictions for sufficiently
large $N$ and $k$. An example is displayed in Fig.~\ref{Fig:scalingkN1024}.

For both, networks of fixed $k$ and networks of fixed $p$, there
are deviations for small and even for intermediate $N$, because the
prediction $r_{\textrm{RMT }}$ was obtained from Random Matrix Theory
that is exact only in the limit $N\rightarrow\infty$, and the finite-size
scaling of $r_{\textrm{RMT}}$ was assumed to resemble the scaling
of the variance of finite matrices. Furthermore, as discussed above,
the numerical estimators of the radius rely on assumptions that are
fulfilled only approximately. For sufficiently large networks, however,
the theoretical prediction agrees well with the numerical data.

Thus there is a gap of size \begin{equation}
g=1-A_{0}-r_{\infty}\end{equation}
 between the non-trivial eigenvalues for large networks and the unit
circle, where \begin{equation}
r_{\infty}:=\lim_{N\rightarrow\infty}r_{\textrm{RMT}}=(1-A_{0})k^{-1/2}.\end{equation}
 This indicates that the stability of the synchronous state in the
model system considered is robust, i.e., sufficiently small perturbations
to the systems dynamics will not alter the stability results. 

Nevertheless, there is an important restriction to these results.
Given a fixed in-degree $k$, the limit $N\rightarrow\infty$ is \textit{not}
described by the theory derived in the previous section, because the
structure of the network considered and thus the structure of the
stability matrices is only well defined if the network is not connected
in the sense that every oscillator has at least one presynaptic oscillator.
However, the probability that at least one oscillator is disconnected
from the remaining network approaches one with increasing network
size. Thus eigenvalue predictions of stability matrices of networks
with fixed in-degree $k$ are only reasonably described for network
sizes that are large, $N\gg1$, but not in the limit $N\rightarrow\infty$.

\subsection{Networks with constant connection probability}

\begin{figure*}[htp!]
\includegraphics[%
  width=16cm]{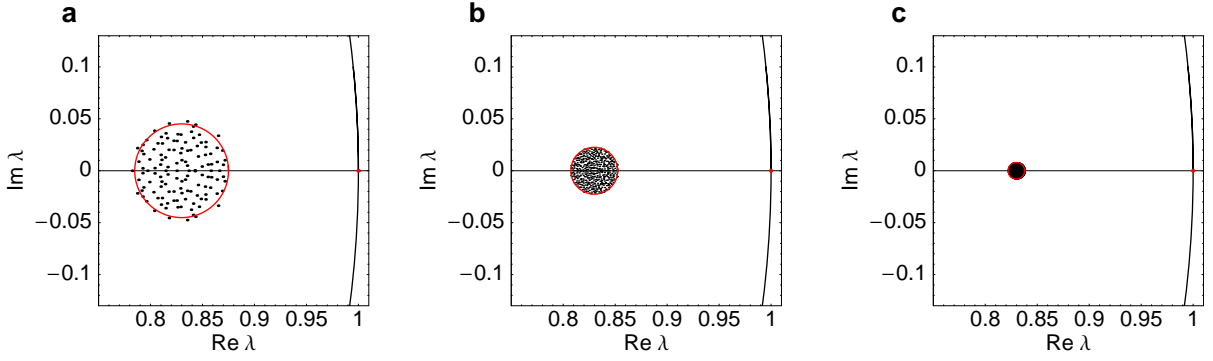}

\caption{Distribution of eigenvalues in the complex plane for networks with
fixed connection probability $p=0.1$ for different network sizes
(a) $N=128$, (b) $N=512$, (c) $N=4096$. The disks are centered
at $A_{0}$ and have radius $r_{\textrm{RMT}}$, the prediction obtained
from Random Matrix Theory. Note that the disk of non-trivial eigenvalues
shrinks towards the point $A_{0}$ in the limit $N\rightarrow\infty$.
The arc through the trivial eigenvalue $z=\lambda_{1}=1$ is a sector
of the unit circle. Parameters of integrate-and-fire oscillators are
$I=1.1$, $\eps=-0.2$, $\tau=0.05$. \label{Fig:EWScircp0.1N}}
\end{figure*}

If we assume that every connection is present with a constant probability
$p$, the network will be connected with probability one in the limit
$N\rightarrow\infty$ because the number of presynaptic oscillators
$k_{i}$ follows a binomial distribution with average $pN$ and standard
deviation $(p(1-p)N)^{1/2}$. In this limit, the radius of the disk
of eigenvalues decreases with increasing network size $N$, see Fig.~\ref{Fig:EWScircp0.1N}. 

\begin{figure}[htp!]
\begin{center}\includegraphics[%
  width=85mm]{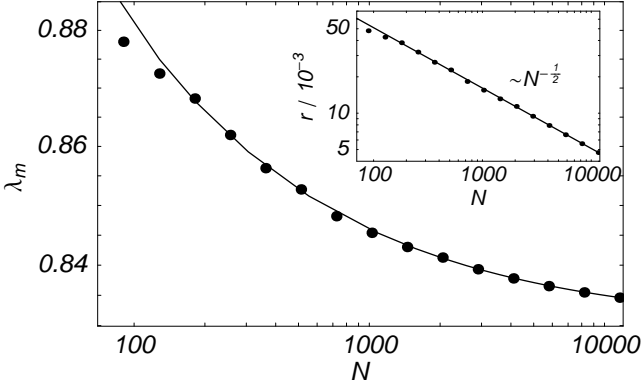}\end{center}

\caption{Maximum non-trivial eigenvalue and the radius of the eigenvalue distribution
for random networks (same parameters as in Fig.~\ref{Fig:EWScircp0.1N}).
Main panel displays the maximal non-trivial eigenvalue $\lambda_{\textrm{m}}\approx A_{0}+r$
as a function of network size $N$. The maximal non-trivial eigenvalue
converges to $A_{0}\approx0.83$ as $N\rightarrow\infty.$ Inset displays
the radius $r$ of the disk of eigenvalues as a function of $N$.
Dots display numerical results based on $r=r_{\textrm{rad}}$ (Eq.~(\ref{eq:rrad})),
lines are the theoretical predictions for both, the radius $r$ and
the maximal non-trivial eigenvalue $\lambda_{\textrm{m}}$.\label{Fig:scalingp0.1N}}
\end{figure}

In order to verify the scaling behavior of the radius of the eigenvalue
disk for large stability matrices $A$, we numerically determined
the eigenvalues $
\lambda_{\textrm{m}}=\textrm{max} \{|\lambda_{i}|\,\,:\,\,|\lambda_{i}|<1\}
$
, see Eq.~(\ref{eq:lambdam}),
that are second largest in absolute value. For sufficiently large
$N$, the theoretical prediction $\lambda_{\textrm{m}}\approx A_{0}+r_{\textrm{RMT}}$
agrees well with the numerical data (Fig.~\ref{Fig:scalingp0.1N}).
The radius approaches zero for large networks such that the eigenvalue
second largest in absolute value converges towards the center $A_{0}$
of the disk. In conclusion, for large networks, all non-trivial eigenvalues
are located near $A_{0}$ and are thus bounded away from the unit
circle. This implies that the speed of synchronization that is determined
by $\lambda_{\textrm{m}}$ increases with increasing network size.
Moreover, the condition necessary for robustness against structural
perturbations of the systems dynamics is satisfied.

\section{Synchronization Speed \newline and Speed Limit }

The existence of bounds on the radius of the eigenvalue distribution
has severe consequences for the synchronization speed of networks
of neural oscillators. Whereas the largest (trivial) eigenvalue $\lambda_{1}=1$
corresponds to the invariant nature of the synchronized periodic orbit,
the second largest eigenvalue $\lambda_{\textrm{m}}$ (Eq.\ \ref{eq:lambdam})
determines the asymptotic speed of synchronization starting from sufficiently
close-by initial conditions. Because the dynamics can locally be approximated
by a linear map, the synchronization of spike times is an exponential.
Thus, denoting $\boldsymbol{\delta}'(t):=\boldsymbol{\delta}(t)-\lim_{s\rightarrow\infty}\boldsymbol{\delta}(s)$,
the distance $\Delta(n):=\max_{i}|\delta'_{i}(nT)|/\max_{i}|\delta'_{i}(0)|$
from the invariant state behaves as\begin{equation}
\Delta(n)\sim\exp(-n/\tau_{\textrm{syn}})\label{eq:delta_tau}\end{equation}
defining a synchronization time $\tau_{\textrm{syn}}$ in units of
the collective period $T$. The speed of synchronization $\tau_{\textrm{syn}}^{-1}$
strongly depends on the parameters. For instance, as might be expected,
synchronization is faster for stronger coupling.

Given the results from Random Matrix Theory derived above, we can
deduce an expression for the synchronization time\begin{equation}
\begin{array}{rcl}
\tau_{\textrm{syn}} & = & -1/\ln(\lambda_{\textrm{m}})\\
 & = & -1/\ln(A_{0}+r_{\textrm{RMT}})\end{array}\label{eq:tausyn}\end{equation}
from the prediction of the second largest eigenvalue $\lambda_{\textrm{m}}\approx A_{0}+r_{\textrm{RMT}}$.
In general, upon increasing the coupling strength $\eps$, the center
$A_{0}$ of the disks of eigenvalues is moved towards zero, as can
be seen from the defining equation (\ref{eq:A_ii}). This means that,
as expected, stronger interaction strengths lead to faster synchronization
of the spiking activity if all other dynamical and network parameters
are kept fixed. There is, however, a speed limit to synchronization
if the in-degree of the network is finite. The speed limit plays a
noticeable role if the typical in-degree $k$ is significantly smaller
than the number $N$ of oscillators in the network.

\begin{figure}[htbp]
\begin{center}\includegraphics[%
  width=80mm]{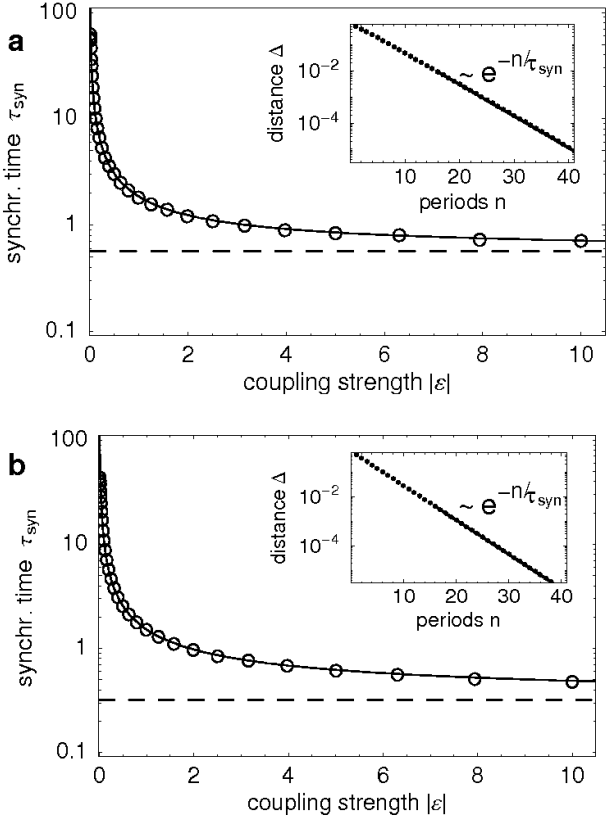}\end{center}

\caption{Asymptotic synchronization time in random networks. (a) Network with
fixed in-degree $k_{i}\equiv k=32$ ($N=1024$, $I=1.1$, $\tau=0.05$,
$\eps_{ij}=\eps/k$ for $j\in\Post(i)$) . The inset shows the distance
$\Delta$ of a perturbation $\boldsymbol{\delta}$ from the synchronous
state versus the number of periods $n$ ($\eps=-0.4$). Its slope
yields the synchronization time $\tau_{\textrm{syn}}$ shown in the
main panel as a function of coupling strength $|\eps|$. Simulation
data ($\bigcirc$), theoretical prediction (------) derived in this
paper, its infinite coupling strength asymptote (--~--~--). (b)
Network of $N=2048$ neural oscillators and connection probability
$p=0.2$ . Other parameters and inset as in (a). Note that although
the typical in-degree is changed drastically from (a) to (b), the
synchronization speed limit is hardly affected. \label{Fig:tausynkp}}
\end{figure}

For networks with constant in-degree (Fig.~\ref{Fig:tausynkp}) ,
the radius of the eigenvalue disk converges to a positive constant
with increasing network size $N$. %
This means that the second largest eigenvalue does not converge to
zero as the coupling increases arbitrarily strong. Thus the asymptotic
synchronization time (\ref{eq:tausyn}) is bounded below by \begin{equation}
\tau_{\textrm{syn }}^{|\eps|\rightarrow\infty}=\frac{2}{\ln k}\left[1+\frac{k}{N\ln(k)}+\mathcal{O}\left(N^{-2}\right)\right]\label{eq:taulimit}\end{equation}
 for large $N$ and thus limited by the network connectivity (cf.\ 
the asymptotes in Fig. \ref{Fig:tausynkp}).

For networks with fixed connection probability $p$, the radius of
the eigenvalue disk does converge to zero with increasing network
size $N$. However, for any finite network with finite number of connections
per oscillator, it has a positive radius and again leads to a speed
limit, see Fig.~\ref{Fig:tausynkp}. Note that in the example displayed
the typical number of connections per oscillator is as large as $k\approx pN\approx409$
but the speed limit is still prevalent. 

\begin{figure}[htbp]
\begin{center}\includegraphics[%
  width=80mm]{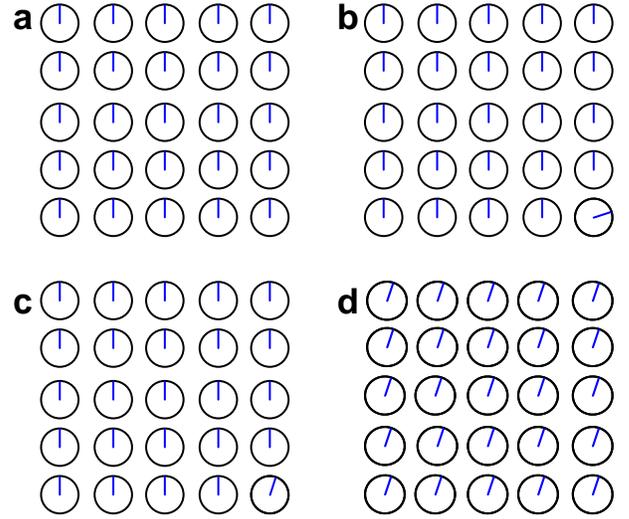}\end{center}

\caption{Schematic illustrating the mechanism of resynchronization in a network
of pulse-coupled neural oscillators. A collection of oscillators (connections
not shown) at specific phases illustrated as {}``time on the clocks''.
(a) Unperturbed, fully synchronous state. (b) One oscillator perturbed
(out of phase). (c) \emph{Purely local} restoring of phases might
seem to be the natural way for resynchronization but it is \emph{not}
\emph{possible} because local averaging of phases implies spreading
of the perturbation such that finally (d) all oscillators of the network
agree on a common phase that does not equal their original one.}

\label{Fig:clocks}
\end{figure}

Can we intuitively understand the speed limit that is enforced by
the topology of the network, parameterized by its typical in-degree?
Consider a large number of neural oscillators connected via a network
of complicated topology. If from the fully synchronous state (Fig.~\ref{Fig:clocks}a)
only one oscillator is perturbed away (Fig.~\ref{Fig:clocks}b) this
constitutes a simple example of resynchronization.%
 One might imagine that all the other oscillators are pulling the
phase of the perturbed one back to their common phase (Fig.~\ref{Fig:clocks}c).
This would explain why, with increasing coupling strengths, synchronization
would be faster -- the stronger the local pulling force, the faster
the local resynchronization. If that was the only mechanism involved,
the network could be resynchronized arbitrarily fast using sufficiently
large coupling strengths. The actual mechanism, however, is non-local.
Because in the linearized dynamics each neural oscillator performs
local averaging, see Eqs.\ (\ref{eq:matrixelements})--(\ref{eq:A0inh}),
of their own phase and those phases of its presynaptic oscillators,
the common phase of the resynchronized state will be globally agreed
on (Fig.~\ref{Fig:clocks}d), i.e. determined by the phases of all
oscillators in the network. Neural oscillators can only interact with
their neighbors, and, due to their pulsed interactions, only at discrete
times once a period. For inhibitory interactions this means that the
time between communication events is bounded below by $\Delta t_{\textrm{interact}}\geq1$,
independent of the delay time $\tau$. At long times, the averaging
has to be performed all over the network, thus restricting the speed
of synchronization.

\section{Conclusion}

We have investigated the dynamics of synchronization in networks of
neural oscillators with complex connection topology. We first described
the stability analysis for the general case and found that the arising
nonlinear and first order mappings have multiple state dependent parts.
As an important consequence standard eigenvalue analysis of the first
order system is not suitable. Using alternative methods, we demonstrated
that the simplest periodic state, the synchronous state in which all
neurons fire periodically at identical times, is stable for inhibitory
coupling, independent of the specific network topology. Second, to
study the speed of synchronization, we derived a subclass of models
for which all parts of the first order stability operators become
degenerate. This class in general requires rise functions of integrate-and-fire
type. Subsequently, we used Random Matrix Theory to analytically predict
the speed of synchronization via the eigenvalue distributions depending
on dynamical and network parameters. Numerical estimates are in excellent
agreement with our theoretical predictions. 

Although the theory used is based on Gaussian (i.e.~fully occupied)
matrices in the limit $N\rightarrow\infty$, our results also hold
for sparse random networks with moderately large finite $N$. Moreover,
it is known that the eigenvalue distribution of the sparse \emph{symmetric}
random matrix ensemble converges towards the eigenvalue distribution
of the Gaussian symmetric ensemble in the limit $k\rightarrow\infty$.
It is not clear whether a similar relation holds for Gaussian and
sparse \emph{asymmetric} ensembles as well. In fact it is an open
question why the the Gaussian ensemble actually describes the synchronization
of sparse random networks even for small $k\approx10^{1}$ rather
than only for $k\rightarrow\infty$.

Our results also indicate that stable synchrony is common to a class
of neural oscillators and not restricted to the specific model considered
here. Moreover, given the expression for the speed of synchronization,
we discovered a speed limit to synchronization on networks that is
controlled by the typical in-degree of each oscillator, i.e. the number
of other oscillators it receives input from. The dependence of the
speed limit on the in-degree is logarithmic such that even for large
in-degree the speed limit is significant. 

The application of Random Matrix Theory in the present study suggests
that it might well be possible to analytically predict dynamical properties
of other systems from their structure, using an ansatz comparable
to ours. Examples of the application of Random Matrix Theory in ecology
are provided in references \cite{May:1976:459,Jirsa:2004:070602}.
They were restricted to the dynamics near fixed points. Due to the
idealization in the model class considered here, it was possible to
analytically predict dynamical aspects near invariant (periodic) solutions
that are not simple fixed points using Random Matrix Theory.

Some straightforward generalizations of possible application include
less simple periodic states like cluster periodic orbits \cite{Ernst:1995:1570,Timme:2002:154105,Timme:2003:377}
or periodic patterns of spikes which occur in the presence of heterogeneity
\cite{Jin:2002:208102,Boergers:2003:509,Denker:2004:074103,Memmesheimer:2006:1}.
More interesting, and certainly more involved possibilities for Random
Matrix Theory applications may arise if the dynamics becomes unstable.
Of particular interest for theoretical neuroscience may be saddle
periodic orbits which imply a high degree of flexibility when switching
between states \cite{Hansel:1993:3470,Rabinovich:2001:068102,Timme:2002:154105,Zumdieck:2004:244103,Ashwin:2005:2035,Ashwin:2005:36}.
Starting from the class of systems considered in the current paper,
the next step into this direction would be to consider orbits that
arise in networks where inhibitory and excitatory recurrent interactions
coexist \cite{vanVreeswijk:1996:1724,vanVreeswijk:1998:1321,Brunel:1999:1621}. 

Our approach is not restricted to the well known Erdos Renyi random
graphs considered here. If other network topologies have to be considered,
we expect that under some additional assumptions, just the associated
random matrix ensemble could be used to describe the linearized dynamics
of such systems. For future investigations of synchronization properties
of networks, scale free and small world \cite{Monasson:1999:555,Farkas:2001:026704}
topologies constitute promising candidates because these networks
might be analytically tractable but nevertheless appear to reflect
important aspects of real world networks.

We thank Michael Denker, Tsampikos Kottos, Peter Müller, Haim Sompolinsky,
Martin Weigt and Annette Zippelius for useful discussions and comments
on this work. Supported in part by the Federal Ministry of Education
and Research (BMBF), Germany, under grant number 01GQ0430 (Bernstein
Center for Computational Neuroscience BCCN Göttingen, Project A2).

%\bibliographystyle{mybib}
%\bibliography{allrefs}

\end{document}